\documentclass[twocolumn,aps,eqsecnum,showkeys,showpacs,superscriptaddress]{revtex4-2}
\usepackage[latin9]{inputenc}
\usepackage{textcomp}
\usepackage{bm}
\usepackage{amsmath}
\usepackage{graphicx}

\begin{document}
%===========================================================================================
% title page
\title{H$_{3}$Se in the $Im\overline{3}m$ Phase: A High-Pressure Superconductor with 
$T_{\text{c}}$ Reaching 200 K at 64 GPa Mediated by Anharmonic Phonons }
\author{Yao Ma}
\affiliation{Department of Applied Physics, School of Sciences, Xi'an University
of Technology, 710054, China}
\author{Mingqi Li}
\affiliation{Department of Applied Physics, School of Sciences, Xi'an University
of Technology, 710054, China}
\author{Wenjia Shi}
\affiliation{Department of Applied Physics, School of Sciences, Xi'an University
of Technology, 710054, China}
\author{Vei Wang}
\affiliation{Department of Applied Physics, School of Sciences, Xi'an University
of Technology, 710054, China}
\author{Pugeng Hou}
\email{pugeng\_hou@neepu.edu.cn}
\affiliation{College of Science, Northeast Electric Power University, Changchun
Road 169, 132012 Jilin, China}
\author{Mi Pang}
\email{pangmi@xaut.edu.cn}
\affiliation{Department of Applied Physics, School of Sciences, Xi'an University
of Technology, 710054, China}
\begin{abstract}
Hydrogen-based compounds have attracted significant attention in recent
years due to the discovery of conventional superconductivity with
high critical temperature under high pressure, rekindling hopes for
searching room temperature superconductor. In this work, we investigated
systematically the vibrational and superconducting properties of H$_{3}$Se
in $Im\overline{3}m$ phase under pressures ranging from 50 to 200
GPa. Our approach combines the stochastic self-consistent harmonic
approximation with first-principles calculations to address effects
from the quantum and anharmonic vibrations of ions. It turns out that
these effects significantly modify the crystal structure, increasing
the inner pressure by about 8 GPa compared to situations where they
are ignored. The phonon spectra suggest that with these effects included,
the crystal can be stabilized at pressures as low as about 61 GPa,
much lower than the previously predicted value of over 100 GPa. Our
calculations also highlight the critical role of quantum and anharmonic
effects on the electron-phonon coupling properties. Neglecting these factors
could result in a substantial overestimation of the superconducting
critical temperature $T_{\text{c}}$, by approximately 25 K at 125 GPa, for example.
With anharmonic phonons, the $T_{\text{c}}$ derived from the 
Migdal-Eliashberg equations, reaches 200 K ($\mu^\star=0.1, \lambda$=4.1) 
as the pressure decreases to 64 GPa, making the crystal a rare high-$T_{\text{c}}$ 
superconductor at moderate pressures.
\end{abstract}
%\pacs{74.70.-b; 63.22.-m}
%\keywords{hydrogen-riched system; superconductivity; high pressure; first principles; stochastic self-consistent harmonic approximation}
\maketitle
%===========================================================================================

\section{Introduction\label{sec:Introduction}}

High critical temperature ($T_{\text{c}}$) superconductivity has
remained to be one of the most attracting yet challenging issues in
physics for decades. Among the progress that have already been made,
the discovery of high $T_{\text{c}}$ in hydrogen-based systems stabilized
by high pressure has drawn considerable research interest in recent
years \citep{gorkov2018,sun2020,flores2020,kong2021,chen2021,zhang2022,minkov2022,sun2024,
eremets2024,cui2019}.
The prosperity of this field originates far from the suspicion of
Neil Ashcroft in several decades ago \citep{ashcroft1968,ashcroft2004},
that pressurized hydrogen or hydrogen dominant alloys could exhibit
superconductivity in high temperature , and then facilitated by the
advancement of high pressure experimental techniques using diamond
anvil cells \citep{boehler2004,flores2020}. The achievement in H$_{3}$S
with a $T_{\text{c}}$ of 203 K at 155 GPa \citep{drozdov2015} and
in LaH$_{10}$ with $T_{\text{c}}$ of 260 K at 180-200 GPa \citep{somayazulu2019}
sparked hope for finding room temperature superconductors. Since then,
the superconductivity of an increasing number hydrogen-based compounds
were investigated both experimentally and theoretically.

In the prediction of superconducting hydrides, computational techniques
based on the density functional theory (DFT) combined with conventional
superconducting theory \citep{cooper1957,allen1975,allen1983} have
been widely applied and play a crucial role, such as in the famous
cases of H$_{3}$S \citep{duan2015}, LaH$_{10}$ \citep{errea2020}
and YH$_{6}$ \citep{peng2017} and many others \citep{gao2010,kim2011,liu2016,li2015,zhong2016,bianco2018,liang2019f,liang2019p,zhang2020,jiang2022,zhao2022,du2023,yang2023,shi2024,chen2024,liu2024}.
During the calculation of superconductivity, the crucial part is the
treatment of ionic vibrations, from which parameters describing the
electron-phonon interaction can be obtained. Conventional methodology
employs the density functional perturbation theory (DFPT) \citep{Baroni2001},
with ions being treated as classical particles. The potential $V(\bm{R})$,
known as the Born-Oppenheimer (BO) potential, is approximated by a
Taylor expansion around ions' equilibrium positions that minimize
$V(\bm{R})$. Typically, this expansion includes up to second-order
terms, referred to as the harmonic approximation. Thus it ignores
the quantum fluctuations of ions and anharmonic nature of the potential.
Recent studies emphasize the crucial impact of these quantum and anharmonic
effects on the structural, vibrational, and notably, the superconducting
properties in many systems especially hydrogen-based compounds \citep{errea2020,borinaga2016,bianco2018,Hou2021,monacelli2021,belli2022,bianco2017,
ribeiro2018,belli2022j,Setty2020,Setty2021,Setty2024}.
For instance, the experimentally found superconductivity in H$_{3}$S
around 200K \citep{drozdov2015} and in LaH$_{10}$ around 250K \citep{drozdov2019,somayazulu2019}
at high pressures can only be well explained with consideration of
the quantum effect and anharmonicity.

Given the high $T_{\text{c}}$ observed in H$_{3}$S, natural interest
arises about H$_{3}$Se, achieved by substituting S with its isoelectronic
mate Se. However, to the best of our knowledge, reports on H$_{3}$Se
are still rare and controversial. Experimental synthesis and metallization
of H$_{3}$Se under high pressure have not yet been achieved. Zhang
et al. \citep{zhang2018} reported the synthesis of H$_{3}$Se in
the Cccm phase at 23 GPa, which was later identified as (H$_{2}$Se)$_{2}$H$_{2}$
\citep{pace2018}. Theoretical predictions by Heil et al. \citep{heil2015}
for H$_{3}$Se in the $Im\overline{3}m$ phase suggested a $T_{\text{c}}$
of 100 K at 190 GPa, and other studies \citep{zhang2015,amsler2019,chang2020}
estimated $T_{\text{c}}$ values ranging from 110 K to 118 K at 200
GPa. Ge et al. \citep{ge2015} reported a $T_{\text{c}}$ of about
163 K, while Flores-Livas et al. \citep{flores2016} suggested approximately
130 K for the same compound at 200 GPa. Additionally, studies \citep{flores2016,zhang2015}
investigated the structural stability of H$_{3}$Se in $Im\overline{3}m$
phase, indicating stability at pressures exceeding 100 and 166 GPa,
respectively. However, none of these calculations explored the influence
of quantum anharmonic effects of ions.

In this work, the vibrational and superconducting properties of H$_{3}$Se
in $Im\overline{3}m$ phase under relatively lower pressures than
the previous studies, ranging from 50 GPa to 200 GPa, are systematically
investigated using first principle calculations. The impacts of the
quantum and anharmonic nature of ionic vibrations are incorporated
through the stochastic self-consistent harmonic approximation (SSCHA).
Contrary to the previous predictions \citep{flores2016,zhang2015},
our calculations indicate that the crystal remains dynamically stable
above approximately 61 GPa, stabilized by the quantum and anharmonic
effects which are essential below pressures previously thought necessary
to achieve stability (109 GPa with harmonic calculations). Significant
modifications are obtained in optical phonons, including softened
bond-bending modes and hardened bond-stretching modes, impacting electron-phonon
coupling properties and suppressing $T_{\text{c}}$ significantly,
by up to approximately 25 K at 125 GPa. 
With anharmonic phonons, the $T_{\text{c}}$ derived from the 
Migdal-Eliashberg equations, reaches 200 K ($\mu^\star=0.1$), and $\lambda$ 
even exceeds 4, as the pressure decreases to 64 GPa.
This study offers valuable
insights for experimentally confirming the superconducting properties
of H$_{3}$Se.

\section{Methods and computational details}

Conventional first principle methods calculate the vibrational properties
of materials using the harmonic approximation, where ions are treated
as classical particles. The potential $V(\bm{R})$ as a function of
the ions' configuration $\bm{R}$, known as the Born-Oppenheimer potential,
is Taylor expanded up to the second order terms around the configuration
in equilibrium $\bm{R}_{0}$ that minimizes $V(\bm{R})$. By diagonalizing
the dynamical matrix 
\begin{equation}
D_{ab}^{h}=\frac{1}{\sqrt{M_{a}M_{b}}}\frac{\partial^{2}V(\bm{R})}{\partial R_{a}\partial R_{b}}\bigg\vert_{\bm{R}_{0}},\label{eq:dmhar}
\end{equation}
one obtains frequencies of the vibrational quanta, i.e., the harmonic
phonons. $a$ and $b$ are combined indices identifying the Cartesian
coordinates of all the ions. $M_{a}$ and $R_{a}$ are mass and position
of atom $a$.

Despite the success of the harmonic approximation, neglecting quantum
and anharmonic effects of ions can lead to significant deviations
in material properties related to ion vibrations, as mentioned in
Sec. \ref{sec:Introduction}. The recently proposed stochastic self-consistent
harmonic approximation (SSCHA) is aimed at addressing this issue \citep{errea2013,errea2014,monacelli2018,bianco2017,monacelli2021}.
Without approximating Born-Oppenheimer potential $V(\bm{R})$, The
SSCHA rigorously incorporates effects of the quantum ionic fluctuations
based on a variational minimization of the free energy, $F[\rho]=\underset{\tilde{\rho}}{\text{min}}\mathcal{F}[\tilde{\rho}],$
with $\mathcal{F}[\tilde{\rho}]=E[\tilde{\rho}]-TS[\tilde{\rho}]$,
keeping all the anharmonic terms. $E[\tilde{\rho}]=\langle K+V(\bm{R})\rangle_{\tilde{\rho}}$
is the total energy with $K$ the kinetic energy operators. $\langle O\rangle_{\tilde{\rho}}=\text{Tr[O\ensuremath{\tilde{\rho}}]}/\text{Tr}[\tilde{\rho}]$
is the quantum average of operator $O$ taken at the trial density
matrix $\tilde{\rho}$ of the system. $T$ is the temperature and
$S[\tilde{\rho}]$ is the entropy. For feasible implementation in
practice, the trial density matrix $\tilde{\rho}=\tilde{\rho}_{\bm{\mathcal{R},\Phi}}$
is constrained to guarantee the distribution probability of ionic
positions to be a Gaussian type and centered at the centroid positions
$\bm{\mathcal{R}}$ with a width $\bm{\Phi}$ decided by the quantum-thermal
fluctuations around them. The SSCHA minimizes $\mathcal{F}[\tilde{\rho}_{\bm{\mathcal{R},\Phi}}]$
as a function of $\bm{\mathcal{R}}$ and $\bm{\Phi}$. During each
step of the minimization, an ensemble of random ionic configurations
in a supercell is extracted from $\tilde{\rho}_{\bm{\mathcal{R},\Phi}}$
and, the total energy and forces of each configuration are calculated
with external \textit{ab initio} code, to obtain the free energy functional
and its derivatives with respect to $\bm{\mathcal{R}}$ and $\text{\ensuremath{\bm{\Phi}}}$.
Using the derivatives, $\bm{\mathcal{R}}$ , $\text{\ensuremath{\bm{\Phi}}}$
and $\tilde{\rho}_{\bm{\mathcal{R},\Phi}}$ are updated to minimize
the free energy based on a preconditioned gradient descent. At the
minimum, the obtained $\bm{\mathcal{R}}_{\text{eq}}$ determine the
averaged ionic positions and the auxiliary force constants $\text{\ensuremath{\bm{\Phi}_{\text{eq}}}}$
represent the fluctuations around these positions.

In the static limits \citep{monacelli2021,Monacelli2021prb,Lihm2021},
phonon frequencies are determined from eigenvalues of the mass rescaled
second order derivatives of the free energy taken at $\bm{\mathcal{R}}_{\text{eq}}$,
\begin{equation}
D_{ab}^{F}=\frac{1}{\sqrt{M_{a}M_{b}}}\frac{\partial^{2}F}{\partial R_{a}\partial R_{b}}\bigg\vert_{\bm{\mathcal{R}}_{\text{eq}}}\label{eq:dmanh}
\end{equation}
known as the free energy Hessian at $\bm{\mathcal{R}}_{\text{eq}}$.
This is the quantum anharmonic analog to the classical harmonic dynamical
matrix $\bm{D}^{h}$ (eq. \ref{eq:dmhar}). Note that the appearance
of negative eigenvalues of $\bm{D}^{F}$ or $\bm{D}^{h}$ (imaginary
phonon frequencies) indicates the structural instability with or without
the quantum and anharmonic effects. Besides the optimization of the
inner cell ionic positions, the SSCHA can also relax the lattice parameters
under an specified pressure, incorporating the quantum effects and
anharmonicity. This is realized by replacing the Born-Oppenheimer
potential with the free energy in computing the stress tensor.

For conciseness, throughout this paper, calculations using the classical
harmonic approximation are referred to as harmonic calculations, while
those employing the SSCHA to include quantum and anharmonic effects
are referred to as anharmonic calculations.

The Eliashberg spectral function 
\begin{equation}
\alpha^{2}F(\omega)=\frac{1}{2\pi N(0)N_{q}}\sum_{\mu\bm{q}}\frac{\gamma_{\mu}(\bm{q})}{\omega_{\mu}(\bm{q})}\delta[\omega-\omega_{\mu}(\bm{q})]\label{eq:a2f}
\end{equation}
is calculated both at the harmonic and anharmonic levels, where $\gamma_{\mu}(\bm{q})$
is the phonon linewidth due to electron-phonon interaction at wave
vector $\bm{q}$ of mode $\mu$. $N(0)$ is the density of states
at the Fermi level. $N_{q}$ is the number of phonon momentum points
used for the BZ sampling. $\omega_{\mu}(\bm{q})$ represent phonon
frequencies, obtained by diagonalizing $\bm{D}^{F}$ or $\bm{D}^{h}$
in the anharmonic or harmonic calculations. The electron-phonon coupling
constant $\lambda$ and the average logarithmic frequency $\omega_{\text{log}}$
can be obtained directly from $\alpha^{2}F(\omega)$ as $\lambda=2\int_{0}^{\infty}\text{d}\omega\left(\alpha^{2}F(\omega)/\omega\right)$
and $\omega_{\text{log}}=\text{exp}\left((2/\lambda)\int_{0}^{\infty}\text{d\ensuremath{\omega}}(\alpha^{2}F(\omega)\text{log}(\omega)/\omega)\right)$.

The superconducting critical temperature $T_{c}$ is determined from
the Allen-Dynes modified McMillan equation \citep{allen1975} 
\begin{equation}
T_{c}=\frac{f_{1}f_{2}\omega_{\text{log }}}{1.2}\text{exp\ensuremath{\left[-\frac{1.04(1+\lambda)}{\lambda-\mu^{*}(1+0.62\lambda)}\right]}},\label{eq:Tc}
\end{equation}
with $\mu^{*}$ the Coulomb pseudopotential. $f_{1}$, $f_{2}$ are
functions of $\lambda$, $\mu^{*}$ and $\alpha^{2}F(\omega)$. For
comparison, $T_{c}$ is also determined from solving the isotropic
Migdal-Eliashberg equations \citep{allen1983} once $\alpha^{2}F(\omega)$
is obtained.

The \textit{ab initio} calculations were performed using the QUANTUM
ESPRESSO (QE) package \citep{giannozzi2009}, employing ultrasoft
pseudopotentials with the Perdew-Burke-Ernzerhof (PBE) parametrization
\citep{Perdew1996} of the exchange-correlation potential. The cutoff
energies for the wave functions and density were set to be 80 Ry and
800 Ry, respectively. Integration over the Brillouin zone in the self-consistent
calculations was carried out using Methfessel-Paxton smearing with
a broadening of 0.01 Ry and a 24\texttimes 24\texttimes 24 $\bm{k}$-point
grid. Harmonic phonon calculations were performed on a 9\texttimes 9\texttimes 9
$\bm{q}$-point grid using DFPT \citep{Baroni2001}. The SSCHA calculations
were conducted on a 3\texttimes 3\texttimes 3 supercell containing
108 atoms, and the resulting anharmonic dynamical matrices were interpolated
to the finer 9\texttimes 9\texttimes 9 grid. The electron-phonon interaction
was evaluated in both the harmonic and anharmonic cases with a Gaussian
smearing of 0.008 Ry.

\section{Results and discussions}

\begin{figure}
\includegraphics[width=8cm]{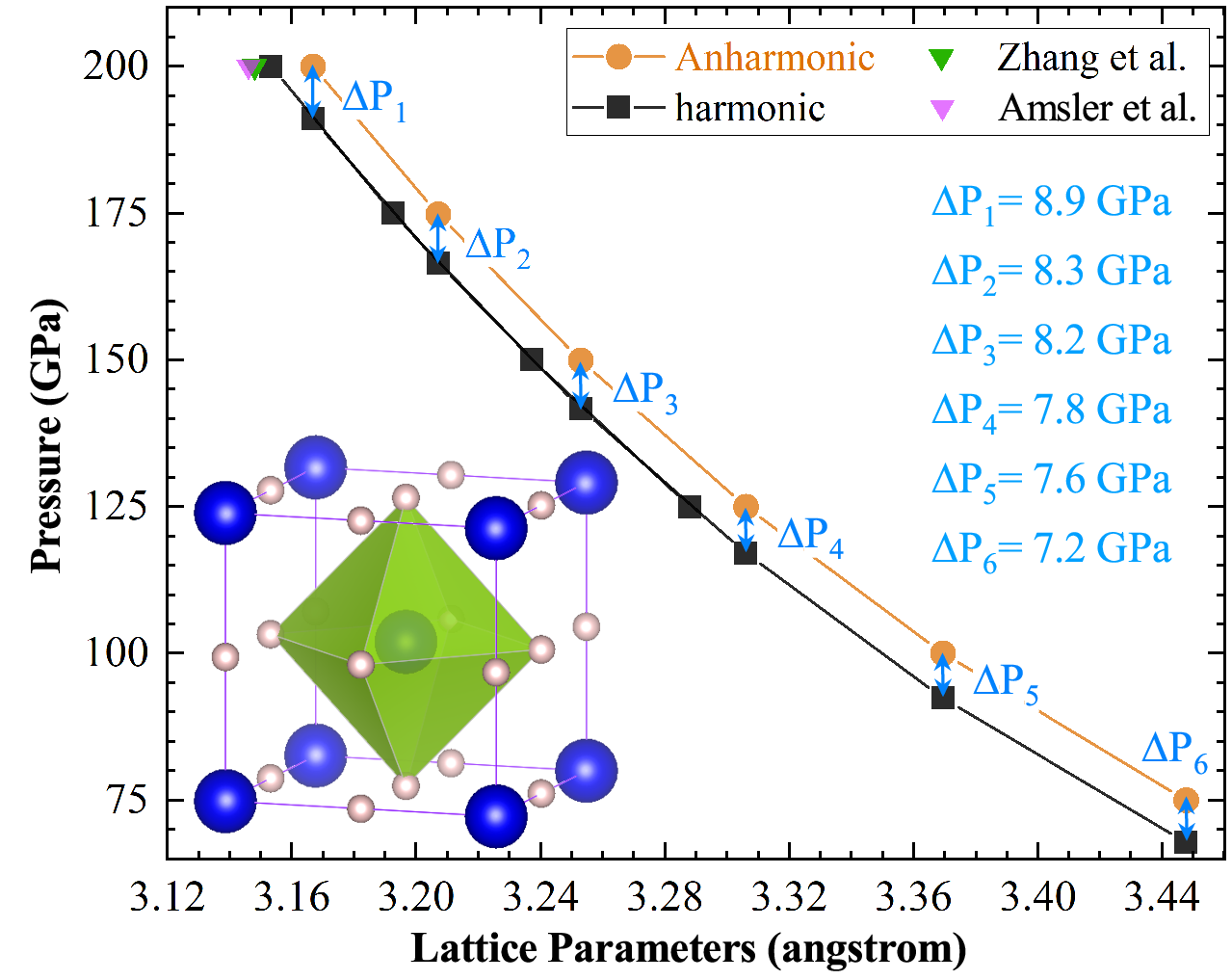} \caption{The crystal structure and the lattice parameter dependence on external
pressure. The crystal structure of $Im\overline{3}m$ H$_{3}$Se is
shown in the corner, with Se atoms represented by blue spheres and
H atoms by pink ones. One of the polyhedra surrounding the central
Se atom is depicted. The dependence of lattice parameter on pressure
is calculated both at the harmonic (black squares) and anharmonic
(orange dots) levels. The results of Zhang et al. \citep{zhang2015}
and Amsler \citep{amsler2019} are also shown by colored triangles.\label{fig:structure}}
\end{figure}

Previous studies \citep{zhang2015,flores2016} suggest that H$_{3}$Se
can be stabilized in the high symmetry $Im\overline{3}m$ phase under
pressures greater than 100 GPa. Herein we focus on this structure
and apply a wider range of pressures to explore the impact of quantum
ionic fluctuations on its structural and vibrational properties. As
shown in the corner of Fig. \ref{fig:structure}, this structure has
the body-centered symmetry, each Se atom has 6 H atoms as the nearest
neighbors locating at the 6 corners of a regular octahedron centered
on the Se atom itself, and the situation is similar for each of the
H atoms. Using both the harmonic and anharmonic calculations, the
structure deformation under external pressure is investigated. The
pressure dependence of the lattice parameter is shown in Fig. \ref{fig:structure}.
Evidently, the quantum and anharmonic effects significantly correct
the stress between ions. For the same lattice parameter, the pressure
obtained from harmonic calculations (referred to as harmonic pressure),
is approximately 8 GPa lower than that from the SSCHA (referred to
as anharmonic pressure). While the slope of pressure with respect
to lattice parameter is hardly affected. This correction in stress
is a commonly seen feature among superhydrides, such as in AlH$_{3}$
\citep{Hou2021} and AlMH$_{6}$ \citep{hou2024}. Our calculation
is consistent with that in the previous studies as shown by triangle
symbols in Fig. \ref{fig:structure}. 

\begin{figure*}
\includegraphics[width=18cm]{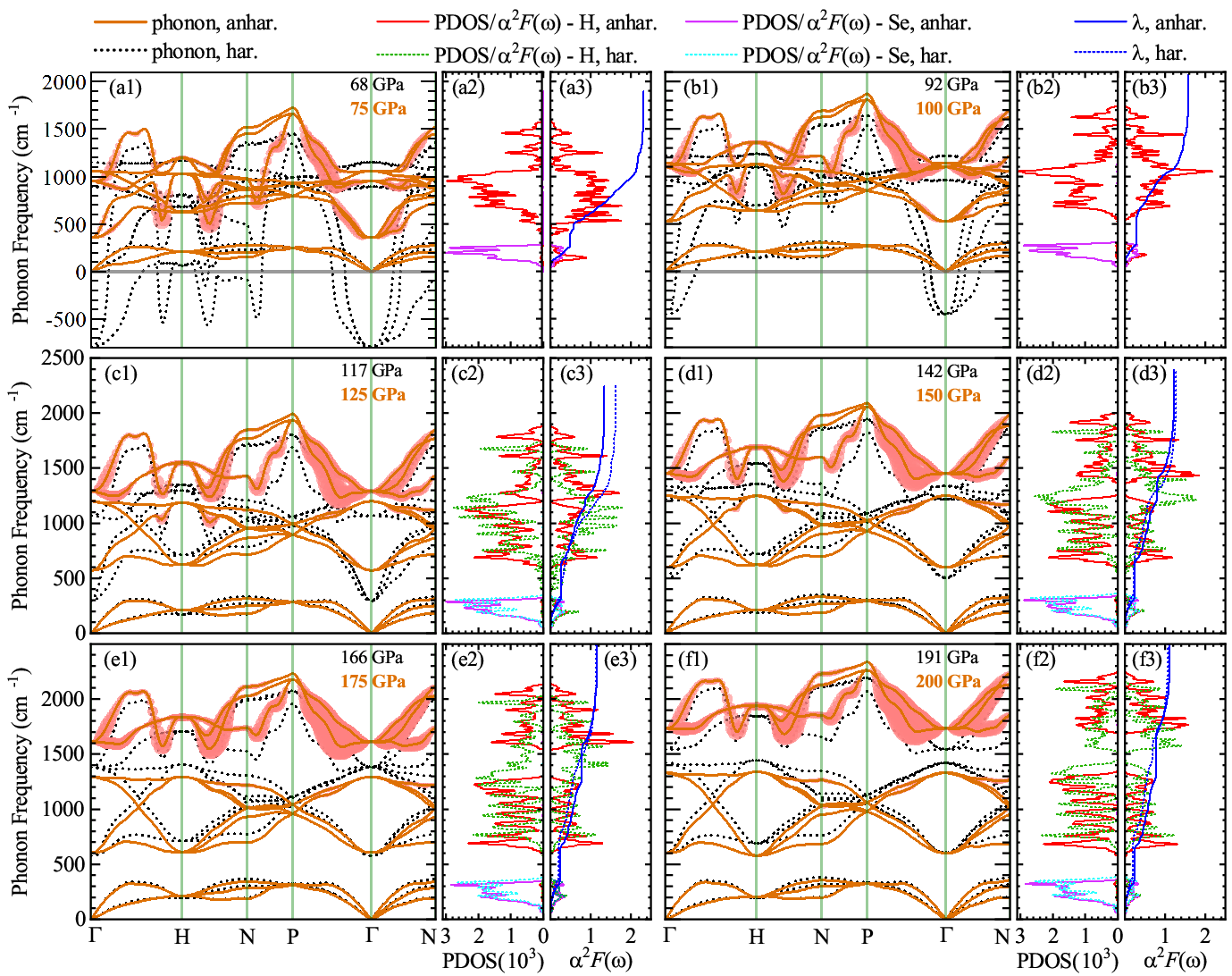} \caption{(a1)$\sim$(f1) The comparison of the harmonic (dotted lines) and
anharmonic (solid lines) phonon spectra of $Im\overline{3}m$ H$_{3}$Se
under pressures from 75 GPa to 200 GPa. The corresponding harmonic
pressures are also marked on each panel with black text. The anharmonic
phonon linewidth of each mode due to electron-phonon interaction is
denoted by the size of the red dots, in arbitrary unit. Zero frequency
is displayed by gray lines. (a2)$\sim$(f2) The projected phonon density
of states (PDOS) and (a3)$\sim$(f3) the projected Eliashberg function
$\alpha^{2}F(\omega)$ corresponding to the spectrum, with solid lines
for anharmonic calculations and dotted lines for harmonic ones, red/green
lines for contribution from H and magenta/cerulean lines for contribution
from Se. The electron-phonon coupling constant $\lambda$ is depicted
with blue lines in (a3)$\sim$(f3). \label{fig:phonon}}
\end{figure*}

In Fig. \ref{fig:phonon} we present the phonon spectra, the corresponding
projected phonon density of states (PDOS) and Eliashberg function
$\alpha^{2}F(\omega)$ of H$_{3}$Se in the $Im\overline{3}m$ phase
across a wide pressure range from 75 GPa to 200 GPa, using both the
harmonic approximation and SSCHA. 

\begin{figure*}
\includegraphics[width=15cm]{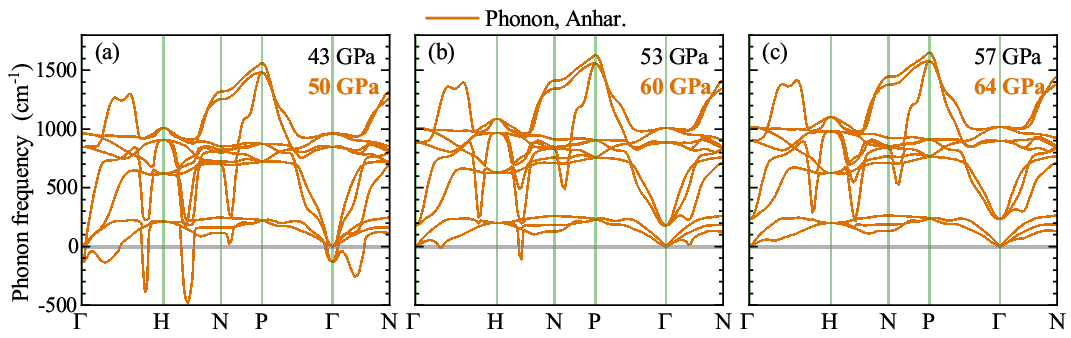} \caption{\label{fig:phonon-2} 
The anharmonic phonon dispersions at (a) 50, (b) 60 and (c) 64 GPa. The harmonic pressures
corresponding to the same crystal are also denoted by black texts on each panel.}
\end{figure*}

The anharmonic pressure (orange text) and the corresponding harmonic
pressure (black text) of the same crystal structure are marked on
each spectrum panel. Similar to the case in AlH$_{3}$ \citep{Hou2021},
under low pressures less than 109 GPa (see panel (a1) and (b1) of
Fig. \ref{fig:phonon}), the harmonic calculation gives negative (actually
imaginary) phonon frequencies, most notably at the $\Gamma$ point,
indicating structural instability. This prediction is roughly consistent
with that in ref. \citep{flores2016}. However, the anharmonic phonon
frequencies obtained from $\bm{D}^{F}$remain positive all through
the 75$\sim$200 GPa range,
highlighting the crucial role of quantum anharmonic effects in stabilizing
the $Im\overline{3}m$ phase of H$_{3}$Se under low pressures.
In order to determine the minimum pressure needed to stabilize the crystal
in the anharmonic case, additional anharmonic calculations are performed
in the low pressure range of 50$\sim$64 GPa (see Fig. \ref{fig:phonon-2}).
The results reveal the emergence of imaginary anharmonic phonons under pressure
around 60 GPa. A interpolation process examing the dependence of
the lowest optical phonon frequency on pressure suggests that the 
minimum pressure needed to stabilize the crystal is approximately
61 GPa.

Within the 75$\sim$200 GPa pressure range, the acoustic phonon branches
are clearly separated from the optical ones and are hardly influenced
by the quantum and anharmonic effects. This is because the acoustic
modes predominantly involve contributions from the heavy selenium
atoms, as evidenced by the phonon partial density of states (PDOS)
in Fig. \ref{fig:phonon} (a2)$\sim$(f2). By contrast, the optical
phonons are significantly affected by quantum effects and anharmonicity
in a way that, the six mid-lying bond-bending branches are roughly
softened while the three high-energy bond-stretching branches are
hardened, except the $\Gamma$ point. Note that under low pressures,
the bond-bending and bond-stretching branches are highly mixed together.
As shown in Fig. \ref{fig:phonon} (a2)$\sim$(f2), hydrogen is the
primary contributor to these optical modes. The anharmonic phonon
linewidth caused by electron-phonon interaction is displayed in Fig.
\ref{fig:phonon} (a1)$\sim$(f1) by the size of the red dots, which
indicate the strong coupling of the bond-stretching phonons with the
electrons. These modes thus contribute significantly to $T_{\text{c}}$.

\begin{figure}
\includegraphics[width=9cm]{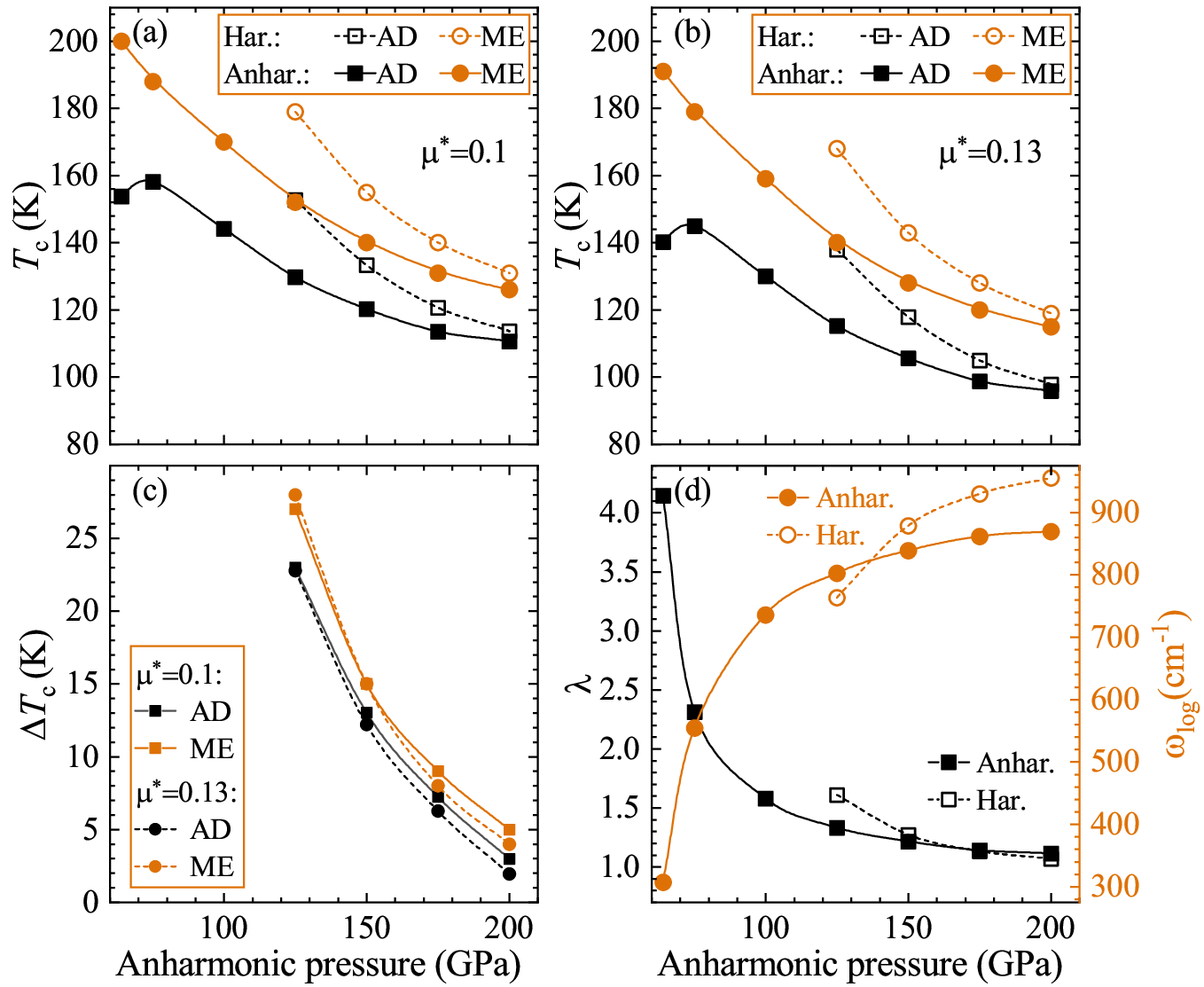} \caption{\label{fig:tc} The calculated $T_{\text{c}}$ of the $Im\overline{3}m$
phase of H$_{3}$Se as a function of the anharmonic pressure. $\mu^{*}$
is chosen as (a) 0.1 and (b) 0.13. $T_{\text{c}}$ calculated with
the harmonic approximation (harmonic $T_{\text{c}}$) are denoted
by hollow symbols, and that calculated by combining the SSCHA and
the electron-phonon coupling matrix (anharmonic $T_{\text{c}}$),
are denoted by solid symbols. Label 'AD' represent results obtained
from the Allen-Dynes modified McMillan equation, and 'ME' indicate
that from the Migdal--Eliashberg equation. (c) The amount by which
harmonic $T_{\text{c}}$ exceeds anharmonic $T_{\text{c}}$, denoted
as $\Delta T_{\text{c}}$. (d) The electron-phonon coupling constant
$\lambda$ (black symbols) and the average logarithmic frequency $\omega_{\text{log}}$
(orange symbols).}
\end{figure}

The strong anharmonic modification on the optical phonons has a deep
impact on the calculated $T_{\text{c}}$. In Fig. \ref{fig:tc}, we
demonstrate $T_{\text{c}}$ obtained from the Allen-Dynes modified
McMillan equation both with the harmonic approximation and by combining
the SSCHA with the electron-phonon coupling matrix elements, which
are called harmonic $T_{\text{c}}$ and anharmonic $T_{\text{c}}$
respectively for conciseness. The Coulomb pseudopotential parameter
$\mu^{*}$ is chosen as the typical values of 0.1 (Fig. \ref{fig:tc}
(a)) and 0.13 (Fig. \ref{fig:tc} (b)). We also solved the Migdal--Eliashberg
equation \citep{allen1983} to obtain more accurate result. The Allen-Dynes
equation underestimates $T_{\text{c}}$ by 16$\sim$50 K compared
with that from the Migdal-Eliashberg equation in the 64$\sim$200 GPa pressure
range both with harmonic and anharmonic calculations. In the strong
electron-phonon coupling case at 75 GPa, the underestimation of $T_{\text{c}}$
by Allen-dynes equation can reach 30 K ($\mu^{*}=0.1$) $\sim $34
K ($\mu^{*}=0.13$), which is 16\%$\sim$19\% of $T_{\text{c}}$.

At 200 GPa, the $T_{\text{c}}$ obtained from Allen-Dynes equation
with $\mu^{\star}=0.1$ in the harmonic calculation is 113.6 K, consistent
with the previous predictions \citep{amsler2019,zhang2015,chang2020}.
For a given $\mu^{*}$, $T_{\text{c}}$ decreases monotonically with
increasing pressure both in the harmonic and anharmonic calculations
and no matter whether the Allen-Dynes or Migdal-Eliashberg equation
is used. The suppression of $T_{\text{c}}$ by pressure is mainly
resulted from the overall hardening of the optical phonon modes imposed
by compression. This can be clearly comprehended from the Allen-Dynes
equation \eqref{eq:Tc}, where $T_{\text{c}}$ increases monotonically
with the electron-phonon coupling constant $\lambda$ within the typical
parameter range. From the formula of $\lambda$ as $\sum_{\mu\bm{q}}\gamma_{\mu}(\bm{q})/[\pi\hbar D(\epsilon_{F})\omega_{\mu}^{2}(\bm{q})]$,
which can be deduced from $\lambda=2\int_{0}^{\infty}\text{d}\omega\left(\alpha^{2}F(\omega)/\omega\right)$
and eq. \eqref{eq:a2f}, it is clear that $\lambda$ decreases, so
as $T_{\text{c}}$, as the vibrations are hardened (phonon frequencies
lifted) and vise versa. $\lambda$ and the average logarithmic frequency
$\omega_{\text{log}}$ are presented in Fig. \ref{fig:tc} (d).

Apparently, $T_{\text{c}}$ is significantly overestimated in the
harmonic approximation due to the neglect of the quantum and anharmonic
effects. The overestimated amount of $T_{\text{c}}$, $\Delta T_{\text{c}}$,
i.e., harmonic $T_{\text{c}}$ minus anharmonic $T_{\text{c}}$, is
presented in Fig. \ref{fig:tc} (c). The overestimation of $T_{\text{c}}$
reaches about 25 K at 125 GPa, and decreases with increasing pressure
to about 4 K at 200 GPa. For further exploration, The projected Eliashberg
spectral function $\alpha^{2}F(\omega)$ and its integral $\lambda(\omega)=\int_{0}^{\omega}2(\alpha^{2}F(\omega')/\omega')d\omega'$,
both for the harmonic and anharmonic cases, are shown in Fig. \ref{fig:phonon}
(a3)$\sim$(f3). The projected $\alpha^{2}F(\omega)$ onto H and Se
can be obtained by calculating the partial contributions from different
types of atoms to the phonon linewidth $\gamma_{\mu}(\bm{q})$ in
eq. \eqref{eq:a2f}. It can be observed from Fig. \ref{fig:phonon}
(a3)$\sim$(f3) that the contribution of the low energy acoustic modes
to $\lambda$ is about 0.3 at 100 GPa and nearly 0.25 at pressures
from 125 to 200 GPa, almost unaffected by the anharmonic effects.
This is quite different from the case of AlH$_{3}$ \citep{Hou2021}
where the contribution of acoustic modes are significantly suppressed
by anharmonicity. The optical phonons, which almost all come from
vibrations of hydrogen atoms, are the main contributors to $\alpha^{2}F(\omega)$.

As shown in Fig. \ref{fig:phonon} (a3)$\sim$(f3) and Fig. \ref{fig:tc}
(d), $\lambda$ is not notably suppressed by anharmonicity at high
pressures from 150 GPa to 200 GPa. $\lambda$ including anharmonicity
is even larger than that with harmonic calculation under 200 GPa (see
blocks in Fig. \ref{fig:tc} (d)), very different from the situations
in AlH$_{3}$ \citep{Hou2021}. This can be interpreted as the result
of a cancellation effect between the softening of bond-bending optical
vibrations and the hardening of bond-stretching ones, both influenced
by anharmonicity. The suppression of $T_{\text{c}}$ by anharmonicity
under pressures from 150 GPa to 200 GPa can be explained by the significant
suppression of the Allen-Dynes average logarithmic phonon frequency
$\omega_{\text{log}}$ by anharmonicity (see Fig. \ref{fig:tc} (d)).
$\omega_{\text{log}}$ is reduced from 930 cm$^{-1}$ to 861 cm$^{-1}$
at 175 GPa and from 955 cm$^{-1}$ to 869 cm$^{-1}$ at 200 GPa by
the quantum and anharmonic effects, which accounts for the decrease
of $T_{\text{c}}$ even so $\lambda$ is slightly lifted when the
anharmonic modification is performed.

Notably, at 64 GPa, $T_{\text{c}}$ calculated from the Migdal-Eliashberg equations
reaches 200 K ($\mu^\star=0.1$), and $\lambda$ even exceeds 4, reveals that 
the $Im\overline{3}m$ H$_{3}$Se is a rare high-$T_{\text{c}}$ superconducting 
hydride at moderate pressures. 
The large value of $\lambda$ here may explain the deviation of the Allen-Dynes $T_{\text{c}}$ 
from the pressure dependence of $T_{\text{c}}$ at higher pressures (as shown in Fig. \ref{fig:tc}(a) 
and (b)), since the Allen-Dynes equation tends to yield inaccurate results in the case of 
strong electron-phonon coupling (large $\lambda $) \cite{xie2022}.

\section{Summary}

To summarize, we investigate systematically the vibrational and superconducting
properties of the $Im\overline{3}m$ phase of selenium hydride H$_{3}$Se
using the stochastic self-consist harmonic approximation (SSCHA) combined
with DFT to incorporate the quantum and anharmonic modifications.
It is found that the crystal structure is significantly influenced
by these modifications. The calculated pressure is about 8 GPa larger
when considering the quantum and anharmonic effects than just calculating
with the traditional harmonic approximation, for the same crystal
structure with lattice parameter from 3.17 to 3.45 \AA. The
phonon spectra are also significantly altered, characterized by the
overall softening of mid-lying optical phonons and hardening of high-energy
ones. Particularly, the phonon spectra imply that the structural instability
predicted by the harmonic calculations below 109 GPa is invalid when
considering the anharmonic modification and, the crystal can actually
remain stable down to approximately 61 GPa. The modification of ionic
vibrational properties exerts a deep effect on the calculated superconducting
critical temperature $T_{\text{c}}$ that, $T_{\text{c}}$ obtained
from the harmonic approximation is significantly suppressed by anharmonicity,
such as from 153 K to 130K at 125 GPa and from 121 K to 113 K at 175 GPa, obtained from the Allen-Dynes
equation. 
Very notably, at 64 GPa, where the crystal maintain dynamically stable facilitated 
by the quantum and anharmonic nature of ions, the $T_{\text{c}}$ calculated
from the Migdal-Eliashberg equations reaches 200 K, reveal that this crystal
is a rare high-$T_{\text{c}}$ superconducting hydride at moderate pressures.
We have analyzed the electron-phonon coupling properties
and calculated parameters, such as phonon linewidth and the Eliashberg
spectral function. These data indicates that almost all corrections
to $T_{\text{c}}$ due to quantum and anharmonic effects stem from
their influence on the vibrations of hydrogen atoms. This highlights
the indispensable role of quantum and anharmonic effects in the estimation
of the ionic vibration related properties in hydrogen selenide, akin
to many other hydrogen-based compounds. Our work would augment the
dataset of the systematic properties of hydrogen-rich compounds, providing
a foundation for further exploration into the essence of superconductivity.

\section{Acknowledgements}

This research was supported by the Natural Science Foundation Project
(Grant No. 20230101280JC) of Jilin Provincial Department of Science
and Technology.

%===========================================================================================
% references
 \bibliographystyle{aipnum4-1}
\bibliography{Ref}

%======================================================================================
\end{document}